\documentclass[a4paper,11pt]{article}
\usepackage{pos}
\usepackage{enumitem}
\usepackage{lineno}[pagewise]
\usepackage{forloop}
\usepackage[para]{footmisc}

\NewDocumentCommand\Nf{mggg}{$N_{\textrm{f}}\text{=}#1\IfNoValueTF{#2}{}{\text{+}#2}\IfNoValueTF{#3}{}{\text{+}#3}\IfNoValueTF{#4}{}{\text{+}#4}$}
\NewDocumentCommand\Nas{mggg}{$N_{\textrm{as}}\text{=}#1\IfNoValueTF{#2}{}{\text{+}#2}\IfNoValueTF{#3}{}{\text{+}#3}\IfNoValueTF{#4}{}{\text{+}#4}$}
\NewDocumentCommand\Nadj{mggg}{$N_{\textrm{adj}}\text{=}#1\IfNoValueTF{#2}{}{\text{+}#2}\IfNoValueTF{#3}{}{\text{+}#3}\IfNoValueTF{#4}{}{\text{+}#4}$}
\NewDocumentCommand\vol{mg}{#1\textsuperscript{3}\IfNoValueTF{#2}{}{$\times$#2}}

\newcounter{linecounter}

\title{Lattice gauge ensembles and data management}

\author[a,\#]{Yasumichi~Aoki}       \affiliation[a]{RIKEN Center for Computational Science (R-CCS), Kobe, 650-0047, Japan}
\author*[b,1]{Ed~Bennett}     \affiliation[b]{Swansea Academy of Advanced Computing, Swansea University, Swansea, United Kingdom}
\author*[c,2]{Ryan~Bignell}     \affiliation[c]{School of Mathematics, Trinity College, Dublin, Ireland}
\author*[d,3]{Kadir~Utku~Can}      \affiliation[d]{CSSM, Department of Physics, The University of Adelaide, Australia}
\author*[e,4]{Takumi~Doi}         \affiliation[e]{Interdisciplinary Theoretical and Mathematical Sciences Program (iTHEMS), RIKEN, Japan}
\author*[f,5]{Steven~Gottlieb}    \affiliation[f]{Department of Physics, Indiana University, IN, USA}
\author*[g,6]{Rajan~Gupta}       \affiliation[g]{Theoretical Division, Los Alamos National Laboratory, NM, USA}
\author*[h,7]{Georg~von Hippel}  \affiliation[h]{PRISMA+ Cluster of Excellence and Institut f\"ur Kernphysik}
\author*[a,8]{Issaku~Kanamori}   
\author*[i,9]{Andrey~Kotov}       \affiliation[i]{J\"ulich Supercomputing Center, Forschungszentrum J\"ulich, Germany}
\author*[j,10,\#]{Giannis~Koutsou} \affiliation[j]{Computation-based Science and Technology Research Center, The Cyprus Institute, Cyprus}
\author*[k,11]{Agostino~Patella}    \affiliation[k]{Humboldt Universit\"at zu Berlin, Institut f\"ur Physik \& IRIS, Germany}
\author*[i,12]{Giovanni Pederiva}   
\author*[l,13]{Christian Schmidt}    \affiliation[l]{Fakult\"at f\"ur Physik, Universit\"at Bielefeld, Germany}
\author*[m,14]{Takeshi Yamazaki}   \affiliation[m]{Institute of Pure and Applied Sciences, University of Tsukuba, Japan}
\author*[n,15]{Yi-Bo~Yang}    \affiliation[n]{CAS Key Laboratory of Theoretical Physics, Institute of Theoretical Physics, Chinese Academy of Sciences, Beijing 100190, China}

\note[1]{For the TELOS collaboration}
\note[2]{For the FASTSUM collaboration}
\note[3]{For the QCDSF collaboration}
\note[4]{For the HAL QCD collaboration}
\note[5]{For the MILC collaboration}
\note[6]{For the Jlab/W\&M/LANL/MIT/Marseille effort}
\note[7]{For CLS}
\note[8]{For the JLQCD collaboration}
\note[9]{For the TWEXT collaboration}
\note[10]{For the ETM collaboration (ETMC)}
\note[11]{For the RC$^\star$ collaboration}
\note[12]{For the OpenLat initiative}
\note[13]{For the HotQCD collaboration}
\note[14]{For the PACS collaboration}
\note[15]{For the CLQCD collaboration}

\note[\#]{Conveners}

\abstract{We summarize the status of lattice QCD ensemble generation
  efforts and their data management characteristics. Namely, these
  proceedings combine the contributions to a dedicated parallel
  session during the 41\textsuperscript{st} International Symposium on
  Lattice Field Theory (Lattice 2024), during which representatives of
  16 lattice QCD collaborations provided details on their simulation
  program, with focus on plans for publication, data management, and
  storage requirements. The parallel session was organized by the
  International Lattice Data Grid (ILDG), following an open call to
  the lattice QCD community for participation in the session.}

\FullConference{The 41st International Symposium on Lattice Field
  Theory (LATTICE2024)\\ 28 July - 3 August 2024\\ Liverpool, UK\\}

\tableofcontents

\begin{document}
\maketitle

\section{Introduction}
The simulation of Quantum Chromodynamics (QCD) via its Euclidean-time,
discrete formulation on a lattice, has been one of the most
compute-intensive applications in scientific computing, consuming
substantial fractions of computer time at leadership HPC facilities
internationally. In particular, the generation of ensembles of gauge
configurations, for multiple values of the QCD parameters such as the
QCD coupling, the quark masses, and the extent of the finite volume,
requires multi-year simulation campaigns, coordinated by multi-member
research collaborations. It is thus common that collaborations store
and reuse the same gauge ensembles for multiple observables of
interest, and in many cases also share the ensembles with researchers
external to the collaboration that generated them.

The purpose of these proceedings is to summarize the available gauge
ensembles generated by various lattice QCD collaborations
internationally, with a focus on the data management practices each
collaboration employs. It follows a parallel session at the
41\textsuperscript{st} International Symposium on Lattice Field Theory
(Lattice 2024), during which 16 collaborations provided status reports
of their simulation efforts, responding to an open call for
participation addressed to the lattice QCD community prior to the
conference. The first such session was held during Lattice 2022 and a
report of the contributions presented during that session can be found
in Ref.~\cite{Bali:2022mlg}.

The sessions are organized by the International Lattice Data Grid
(ILDG) as in-person continuations of a series of online workshops with
the intention of collecting and gaining an overview of ongoing
simulation efforts and the evolving needs of the lattice community in
terms of data storage and management.  The ILDG was setup in the early
2000s~\cite{Davies:2002mu,Yoshie:2008aw,Maynard:2009szr,Beckett:2009cb}
by the lattice community, which realized early on the value in
standardizing data management practices across the field. ILDG is
organized as a federation of autonomous \textit{regional grids},
within a single Virtual Organization~\cite{ildg-organization}. It
standardizes interfaces for the services, which are to be operated by
each regional grid, such as storage and a searchable metadata catalog,
so that the regional services are interoperable. Within ILDG, working
groups specify community-wide agreed metadata schemas
(QCDml)~\cite{Coddington:2007gz} to concisely mark-up the gauge
configurations and develop relevant middleware tools for facilitating
the use of ILDG services. The middleware and metadata specifications
developed by ILDG adhere to most of the FAIR (Findable, Accessible,
Interoperable, Reusable) principles~\cite{Wilkinson2016}. A summary of
recent developments in ILDG, referred to as ILDG 2.0, was presented
during the same session and can be found in a separate proceedings
contribution~\cite{ILDG2}.

In the remainder of these proceedings, we present the status of
ensemble generation of the 15 collaborations that contributed to this
proceedings contribution. The ensembles reported were generated using
\Nf{2}{1}, \Nf{2}{1}{1}, and \Nf{1}{1}{1}{1} sea-quark flavors with
various fermion discretizations. The contributors were asked to
specify whether their data are public or if they plan on making them
public, their interest in using ILDG services and tools for that
purpose, as well as some overall information regarding storage
requirements. This information is collected in a table and summary
section that follows the individual contributions.

\section{Contributions}
The contributions from each collaboration follow, in the order
presented during the parallel session. The original presentations can
be found on the conference website~\cite{parallel-session}.

\subsection{CLQCD}

The CLQCD collaboration focuses on the first principles QCD study of
the spectrum of exotic hadrons, parton structure of the nucleon, and
other traditional hadron, N-point correlation functions related to
high accuracy tests of the standard model, and also QCD in extreme
conditions. For this purpose, CLQCD generated a set of configurations
of \Nf{2}{1} fermions using the tadpole improved clover fermion action
with stout smearing and the tadpole improved Symanzik gauge action, at
5 values of the lattice spacing $\in[0.05,0.11]$~fm, several pion
masses $\in[120, 350]$~MeV, several volumes with
$m_{\pi}L\in[2.6,5.8]$~fm, and several temperatures
$\in(0,464]$~MeV. In addition, there are several anisotropic ensembles
  using clover fermions with $\xi=5$ at different lattice spacing,
  pion mass, volume and flavors which concentrate on the high
  precision studies related to glueball properties. For the gauge
  generation the Chroma package~\cite{Edwards:2004sx} with
  QUDA~\cite{Clark:2009wm,Babich:2011np,Clark:2016rdz} is used at
  present, and we plan to switch to PyQUDA~\cite{Jiang:2024lto} with
  QUDA in the near future. Currently the CLQCD collaboration has
  $\sim$20 ensembles (one ensemble corresponds to one point in the
  space lattice spacing-spatial volume-pion mass-temperature), which
  occupy $\sim 100$ TB of disk space. Configurations are stored in the
  SCIDAC format. Possible collaborations on the analysis of the
  correlation functions are welcome and CLQCD is preparing the
  hardware and software for the data sharing service.

\subsection{Jlab/W\&M/LANL/MIT/Marseille}
The Jlab/W\&M/LANL/MIT/Marseille collaborations are generating
ensembles of (2+1)-flavor QCD including the Sheikholeslami-Wohlert
(SW) term in the fermion action (the Wilson-clover action) and a
tree-level tadpole improved Symanzik gauge action. To smooth the gauge
fields, one iteration of four-dimensional stout smearing, with weight
$\rho = 0.125$ for the staples, is used in the rational hybrid Monte
Carlo (RHMC) algorithm.  After stout smearing, the tadpole improved
tree-level clover coefficient is found to be very close to the
non-perturbative value as checked by using the Schr\"odinger
functional method for determining the clover coefficient
non-perturbatively. The two light flavors ($u$ and $d$ quarks) are
taken to be degenerate and ensembles are generated with $M_\pi \approx
270,\ 220,\ 170$ and 130 MeV and with lattice spacings $a \approx
0.12,\ 0.091,\ 0.071$ and 0.056 fm. The strange quark mass is tuned
close to its physical value by requiring the ratio $(2M_{K^+}^2 -
M_{\pi^+}^2)/M_{\Omega^-}$ takes on its physical value 0.1678, which
is independent of the light quark masses to lowest order in
$\chi$PT. Current analyses of the 13 ensembles generated so far,
however, show that the kaon mass varies between 490--570 MeV,
necessitating all analyses be done with $M_\pi$ and $M_K$ as
independent parameters.  All eleven larger volume ensembles have
$M_\pi L > 4$.  The setting of the lattice scale $a$ and quark masses
$m_l$ and $m_s$ is being done using a global analysis of the octet and
decuplet baryon spectrum, pion and kaon masses and decay constants and
the flow parameters $w_0$ (or $t_0$). Details of the ensembles and
their characterization, based on the results of this analysis of the
spectral quantities, will be presented by April 2025. These ensembles
constitute between 10K-20K thermalized trajectories, with every second
one stored at the Oak Ridge Leadership Computing Facility, and
constitute about 3.5 PB of data. They are available to USQCD mambers
for non-competitive calculations.

\subsection{HotQCD}
The HotQCD collaboration generates finite temperature ensembles, using
highly improved staggered quarks with a tree level improved Symanzik
gauge action (HISQ/tree). The majority of the ensembles are with
(2+1)-flavor of dynamical quark at the physical point, i.e. two
degenerate light quarks ($m_l$) and one heavier strange quark, with a
fixed ratio of $m_l/m_s=1/27$. Here we have $\mathcal{O}(10)$
different temperatures and lattices with aspect ratio
$N_\sigma/N_\tau=4$ and $N_\tau=8,12,16$. These ensembles have been
produced with high statistics for the Taylor expansion of the pressure
at finite baryon chemical potential \cite{Bollweg:2021vqf,
  Bollweg:2022fqq}. In addition we have large lattices with
$m_l/m_s=1/5, 1/20$, $N_\sigma=80,96,127$, $N_\tau=20,22,24,28,32,36$
and $T=195, 220, 251, 293$ MeV, for the calculation of spectral
functions and transport coefficients
\cite{Altenkort:2023eav,Altenkort:2023oms}. Finally we have ensembles
at lighter then physical quark masses: $m_l/m_s=1/40,1/60,1/80,1/160$,
to study the chiral phase transition \cite{Ding:2024sux,
  HotQCD:2019xnw}. Here we increase the aspect ratio with decreasing
quark mass in order to keep the finite size effects under
control. There is also a limited number of ensembles with three and 5
degenerate HISQ fermions, generated for a study of the Columbia plot
\cite{Dini:2021hug,Karsch:2022yka}.  The lattices have been generated
using the SIMULATeQCD code \cite{HotQCD:2023ghu}, using a Rational
Hybrid Monte Carlo (RHMC) with a 3-level leap-frog or Omelyan
integrator. The step sizes of light, strange and gauge force
integrators are tuned to a meet an acceptance of ~70\%. The trajectory
is usually between 0.5 and 1.0 time units. The configurations are
stored in a compressed fp32 format and sum up to $~1,500$ TB. We plan
to upload these ensembles to ILDG 2.0, which is however subject to the
available storage space.


\subsection{FASTSUM}
The FASTSUM collaboration use \Nf{2}{1} flavor anisotropic gauge
ensembles in the fixed-scale approach to study the behavior of QCD as
a function of temperature in hadronic and plasma phases. Specifically
we have considered the behavior of hadronic states including light,
strange, charm and bottom quarks, the electrical conductivity of QCD
matter, the interquark potential and properties of the chiral
transition.
FASTSUM gauge fields utilise an $\mathcal{O}\left(a^2\right)$ improved
Symanzik gauge action and an $\mathcal{O}\left(a\right)$ improved
spatially stout-smeared Wilson-Clover action following the parameter
tuning and zero-temperature ensembles of the Hadron Spectrum
collaboration~\cite{Edwards:2008ja,HadronSpectrum:2008xlg}. ``Generation
2'' ensembles were generated using the Chroma~\cite{Edwards:2004sx}
software suite while the newer ``Generation 2L'' used a
modfication~\cite{glesaaen_jonas_rylund_2018_2216355} of the {\sc
  openQCD}~\cite{Luscher:2012av,openqcd} package which introduces
stout-link smearing and anisotropic actions. Generation 2 and 2L have
an anisotropy $\xi = a_s/a_\tau \sim 3.5$ with $a_s\sim 0.12$ fm, $N_s
= 24$ or $32$ and a wide range of $N_\tau$ corresponding to $T
\in[44,760]$ MeV. 
mainly in their quark mass.  Full details of these ensembles may be
found in Refs.~\cite{Aarts:2014nba,Aarts:2020vyb}. We are in the
process of production for ``Generation 3'' - a parameter set similar
to ``Generation 2'' but with twice the anistropy $\xi \sim 7$ - using
{\sc openQCD-FASTSUM}~\cite{glesaaen_jonas_rylund_2018_2216355}.
We maintain a centralised metadata repository detailing (among other
information) who was responsible for each run, on which machine that
run was produced and where copies may be found. The gauge fields are
redundantly stored on two (well-separated) storage servers managed by
Swansea University in the openQCD format. The ``Generation 2''
ensembles are publically available~\cite{aarts_2024_8403827} while
other ensembles will be available after an embargo period. We
anticipate making ensembles available through the next incarnation of
the ILDG with supplementary information also available on
Zenodo~\cite{zenodo,aarts_2024_8403827}.

\subsection{TELOS}

The TELOS collaboration performs
\textbf{T}heoretical \textbf{E}xplorations on the \textbf{L}attice
with \textbf{O}rthogonal and \textbf{S}ymplectic groups.
Problems of interest focus on physics beyond the Standard Model,
in particular composite Higgs models.
Our work to date has made use of
the Wilson gauge action and Wilson fermion action.
Our ensembles include studies of
the Sp(4) theory with two fundamental fermion flavours
(\Nf{2})~\cite{Bennett:2019jzz}
(five values of $\beta\in[6.9,7.5]$,
$V \le 48\times42^{3}$,
$m_{\mathrm{PS}}/m_{\mathrm{V}} \gtrsim 0.407(16)$),
the Sp(4) theory with three antisymmetric fermion flavours
(\Nas{3})~\cite{Hsiao:2022gju},
(six values of $\beta\in[6.6,6.9]$,
$V \le 54\times36^{3}$,
$m_{\mathrm{ps}}/m_{\mathrm{v}} \gtrsim 0.7954(44)$);
and the Sp(4) theory with \Nf{2} and \Nas{3}~\cite{Bennett:2022yfa},
(three values of $\beta\in[6.45,6.5]$,
$V \le 56\times36^{3}$,
$m_{\mathrm{PS}}/m_{\mathrm{V}} \gtrsim 0.8768(30)$,
$m_{\mathrm{ps}}/m_{\mathrm{v}} \gtrsim 0.9022(27)$).
In the latter two cases,
the topological charge becomes slow running at small $m_{\mathrm{as}}$,
and at larger $\beta$.
We do not retain our pure gauge ensembles,
used for studies of the large-$N$ limit of Sp(2$N$)~\cite{Bennett:2023qwx,Bennett:2022gdz,Bennett:2020qtj},
since the costs of storage and data transfer are
higher than those of regenerating the ensembles.

Additionally,
we present ensembles generated by a subset of the collaboration,
with applications to conformal and near-conformal dynamics,
and to potential Walking Technicolor theories,
again using the Wilson gauge and Wilson fermion actions~\cite{Athenodorou:2024rba}.
Specifically,
these are SU(2) with one adjoint flavour (\Nadj{1})
(seven values of $\beta\in[2.05,2.4]$,
$V \le 96\times48^{3}$,
$m_{2^{+}_{s}} \gtrsim 0.28$),
and \Nadj{2}
($V \le 128\times64^{3}$,
$m_{2^{+}_{s}} \gtrsim 0.47$).
In the former case,
the majority of ensembles show ergodic topology,
with $\beta=2.4$ being marginal;
in the latter,
at large volumes we see significant topological freezing.

Ensembles are generated using
HiRep~\cite{Bennett:2019cxd,DelDebbio:2008zf}
and Grid~\cite{Bennett:2023gbe,Yamaguchi:2022feu}.
The above ensembles will be made available
as soon as the infrastructure is in place to do so.

We are in the process of generating ensembles
for Sp(4) \Nf{2}
and for SU(2) \Nf{1,2}
with Möbius domain wall fermions,
and for SU(2) \Nf{1,2}
with Wilson fermions
and additional Pauli--Villars fields,
which we aim to make available concurrently with the corresponding papers.

\subsection{HAL QCD}
The Hadrons to Atomic nuclei from Lattice QCD (HAL QCD) collaboration
studies multi-hadron systems on the lattice and determines the
interactions among hadrons, which are crucial quantities to construct
a bridge between QCD and nuclear physics as well as astrophysics.
In order to perform the corresponding lattice calculations, it is
necessary to prepare gauge configurations at the physical point with
large physical volume(s) and a large number of ensembles, since the
hadron interactions (whose ranges are typically $\sim 1/m_\pi$) are
sensitive to quark masses, and statistical fluctuations of
multi-hadron systems are known to be large.

For this purpose, we generate a new set of configurations with
\Nf{2}{1} non-perturbatively improved Wilson-clover fermions with
stout smearing and the Iwasaki gauge action on a $96^4$
lattice~\cite{Aoyama:2024cko}.  Utilizing the simulation parameters
taken from those used in the PACS10
configurations~\cite{Ishikawa:2018jee,PACS:2019ofv}, configurations
were generated on the supercomputer Fugaku for 8,000 trajectories at a
single lattice spacing, $a = 0.084$~fm, and at the physical point,
$m_\pi = 137$~MeV, where we employ Hybrid Monte Carlo (HMC) with the
domain-decomposed HMC algorithm and mass preconditioning for up and
down quarks and with the rational HMC algorithm for the strange quark.
The gauge configurations are stored for every fifth trajectory and
thus amount to 1,600 configurations ($\sim 80$ TB) in total. The set
is named ``HAL-conf-2023''~\cite{Aoyama:2024cko}.
We also performed Coulomb gauge fixing for the obtained
configurations.  Utilizing rotational symmetry, we have $1,600 \times
4$ (rotations) = 6,400 gauge-fixed configurations, which amount to
$\sim 320$ TB.  We plan to to make these configurations public through
JLDG/ILDG in the future, after performing the analyses which is
currently on-going.

\subsection{TWEXT}
The TWEXT (Twisted Wilson @ Extreme conditions) collaboration studies
the properties of QCD at high temperature using Wilson Twisted Mass
fermions. Problems under investigation include chiral properties of
QCD, in particular the behavior of QCD around the chiral phase
transition and its scaling window~\cite{Kotov:2021rah}, topological
properties of QCD and QCD axion~\cite{Kotov:2021ujj}, hadron masses,
symmetries of QCD and others. For this purpose, TWEXT generated a set
of configurations for \Nf{2}{1}{1} fermions at the physical pion mass
and also uses older configurations with heavier pion
mass. Configurations with the physical pion mass have three lattice
spacings $a\in(0.057,0.080)$~fm and cover a wide range of temperatures
from $\sim120$ MeV to $\sim900$ MeV. It allows the TWEXT collaboration
to perform the continuum extrapolation for quantities of interest in
this temperature range. For the generation the tmLQCD software
package~\cite{Jansen:2009xp,Deuzeman:2013xaa,Abdel-Rehim:2013wba} is
used and the parameters of the ensembles were taken from the zero
temperature simulations of the ETM
collaboration~\cite{Alexandrou:2018egz}. Currently, the TWEXT
collaboration has 80 ensembles (one ensemble corresponds to one point
in the space temperature-pion mass-lattice spacing), which occupy
$\sim80$ TB of disk space. Configurations are stored in the ILDG
format. Possible collaborations are welcome and TWEXT plans to make
configurations public/use ILDG in the future, after performing the
ongoing analysis.

\subsection{QCDSF}
The main focus of the QCDSF collaboration is hadron spectrum and
structure at zero temperature.
Our ensembles are generated using the Symanzik improved gauge action
and Stout Link Non-perturbative Clover (SLiNC) fermion action, for
which the link variables appearing in the Dirac term are stout
smeared, while the links in the clover term are
not~\cite{Cundy:2009yy}. The clover coefficient is determined
non-perturbatively.
Our most recent set of ensembles are $2+1$-flavour, which cover pion
masses ranging $m_\pi^{phys} \lesssim m_\pi \lesssim 470 \; {\rm
  MeV}$, and 5 lattice spacings in between $a = 0.052 - 0.082 \; {\rm
  fm}$ (inclusive).
In total, there are 22 ensembles available and an additional 2 at
almost-physical pion mass still being generated. A recent listing of
available ensembles can be found in~\cite{QCDSFUKQCDCSSM:2023qlx}.
Our approach to the physical point follows the $\bar m^R = (2m_\ell^R
+ m_s^R)/3 = const$ line~\cite{Bietenholz:2011qq}, i.e. we start from
the SU(3) symmetric point where the renormalized masses of strange
$(m_s^R)$ and light quarks $(m_\ell^R)$ are equal to each other,
$m_s^R = m_\ell^R = \bar m^R/3$ and we increase $m_s^R$ as $m_\ell^R$
decreases.
The BQCD software suite~\cite{Haar:2017ubh} is used to generate the
ensembles, utilising the hybrid Monte Carlo (HMC) and rational HMC
algorithms.
All of our gauge configurations are stored in the ILDG format with
metadata compliant with the ILDG scheme. We have made use of ILDG
storage systems before, where the hub at the CSSM, Adelaide served as
one of the regional grids, and some older configurations are still
stored in the ILDG servers.
The QCDSF collaboration kindly asks any prospective users to contact
the collaboration before utilizing these configurations for their
projects. We plan to make newly generated ensembles available upon
request through ILDG, pending the collaboration's confirmation. QCDSF
is open to new collaborative projects.


\subsection{OpenLat}
The Open Lattice Initiative is generating QCD ensembles using
Stabilized Wilson Fermions (SWF) with \Nf{2}{1}\cite{Francis:2022hyr},
based on the exponentially improved Dirac Wilson fermion formulation.
The code used for the generation is OpenQCD\cite{openqcd}.  These
ensembles are intended to be general-purpose datasets, shared with the
broader LQCD community in line with the principles of open
science. Lines of constant physics are defined through the $\phi_4$
parameter setting at a fixed trace M. The initial tuning was performed
at the $\mathrm{SU}(3)$ flavor symmetric point.  Additionally, a set
of more chiral trajectories has been put into production, identified
by four different pion masses, aiming to reach the physical point.

The generation has been divided into three stages, each to be released
with a publication. The first stage, which is nearly complete, focuses
on the flavor symmetric point; the second on $m_\pi=300,~200$~MeV; and
the third on the physical point. For each pion mass, there are four
lattice spacings with periodic boundary conditions
($a=0.12,~0.094,~0.077,~0.064$~fm) and one ensemble with open boundary
conditions at a finer lattice spacing of $a=0.054$~fm.  The volumes
have been chosen to always ensure that $m_\pi L > 4$.  The
collaboration is strongly in favor of uploading configurations to the
ILDG, as it aligns with our core principles. The files are already
stored in the ILDG format and a preliminary upload of a few
configurations has been done before the LATTICE24 conference. Our
ensembles are publicly available for use after the publication of the
corresponding stage has been completed. The total storage required is
expected to be 0.5~PB.

\subsection{RC$^\star$}
The RC$^\star$ Collaboration is generating QCD and QCD+QED ensembles
with 3 or 4 flavours. We employ C-periodic (\textit{aka} C$^\star$)
boundary conditions in space, which allow to have a local and
gauge-invariant formulation of QED in finite
volume~\cite{Lucini:2015hfa}. For the SU($3$) gauge field we use the
L\"uscher-Weisz action, while for the compact U($1$) field we use the
Wilson action. The Wilson-clover discretization of the Dirac operator
is used, which includes two Sheikholeslami-Wohlert (SW) terms for the
coupling to the SU($3$) and U($1$) field-strength tensors. The
coefficients are chosen in such a why that $O(a \alpha_{\text{em}}^0)$
improvement is guaranteed non-perturbatively in
$\alpha_{\text{s}}$. All ensembles generated so far have a lattice
spacing of about $0.05\text{ fm}$, a heavier-than-physical charged
pion with mass between $360\text{ MeV}$ and $500\text{ MeV}$. We plan
to move towards lighter pions in the near future. The spatial lattice
extent $L$ ranges between $2.9M_\pi^{-1}$ and $6.7M_\pi^{-1}$. Three
values of the fine-structure constant $\alpha_{\text{em}}$ have been
considered so far besides zero, i.e. its physical value, $2.7$ and
$5.5$ times its physical value. In all cases, the $\beta$ for the
U($1$) is smaller than $0.035$, ensuring that the QED sector is deep
in the perturbative regime and away from the bulk phase transition of
the compact U($1$) model. The configurations are generated with the
publicly-available \texttt{openQ*D}
code~\cite{Campos:2019kgw,10261_173334}, which is based on
openQCD-1.6. Some of the ensembles as well as the used algorithms are
described in~\cite{RCstar:2022yjz}. At the moment we have 80Tb of data
stored on the machines where they have been generated. This includes
configurations, input and log files, histories of various reweigthing
factors (to correct for the rational approximation and for occasional
change of sign of the fermionic Pfaffian) and basic observables
(gradient-flow observables and meson correlators). The RC$^\star$
Collaboration has worked closely with the ILDG metadata working group
to extend the XML schema in order to accommodate metadata for QCD+QED
configurations with various finite-volume formulation of QED. We plan
to generate the XML metadata and make our ensembles publicly available
via ILDG as soon as possible.


\subsection{ETMC}
The ETM collaboration focuses on hadron spectroscopy, hadron
structure, and flavor physics at zero temperature. Ensembles employ
the twisted mass formulation, realizing $\mathcal{O}(a)$-improvement
by tuning to maximal twist, and include a clover term to further
reduce the size of lattice artifacts. The Iwasaki gauge action is
used. The main simulation effort is for the generation of ensembles
with degenerate up- and down-, strange- and charm-quarks
(\Nf{2}{1}{1}) with lattice spacing ranging between $0.049$ and
$0.091$~fm. $M_\pi\cdot L$ varies from $2.5$ up to ${\sim}5.5$. At the
time of writing, 24 ensembles are available or in the process of being
generated, with 8 of these at approximately physical values of the
quark masses. For a recent listing of the ensembles,
see~\cite{ETMCPoster:2024}. Simulations are performed using the Hybrid
Monte Carlo (HMC) algorithm implemented in the tmLQCD software
package~\cite{Jansen:2009xp,Deuzeman:2013xaa,Abdel-Rehim:2013wba}. See
Ref.~\cite{Alexandrou:2018egz} for details on the simulation program,
including the parameter tuning. The
DD-$\alpha$AMG~\cite{Frommer:2013fsa,Alexandrou:2016izb} multigrid
iterative solver is employed for the most poorly conditioned monomials
in the light sector while mixed-precision CG is used
elsewhere. Multi-shift CG is used together with shift-by-shift
refinement using DD-$\alpha$AMG~\cite{Alexandrou:2018wiv} for a number
of small shifts for the heavy sector. tmLQCD has interfaces to
QPhiX~\cite{Joo:2013lwm} and QUDA~\cite{Clark:2009wm,Babich:2011np}.
tmLQCD automatically writes gauge configurations in the ILDG format,
with meta-data including creation date, target simulation parameters,
and the plaquette. ETMC policy is to make ensembles publicly
available after a grace period. Older \Nf{2} and \Nf{2}{1}{1}
ensembles~\cite{Baron:2010bv,EuropeanTwistedMass:2010voq,ETM:2009ztk}
have made use of ILDG storage elements. The current ensembles are
available upon request and the collaboration intends to use ILDG in
the near future. For these ensembles, we expect storage requirements
to reach 3~PB.

\subsection{JLQCD}
The JLQCD collaboration uses a tree-level Symanzik improved gauge
action and M\"obius domain wall fermions with the scale factor 2 and 3
levels of stout smearing, of which details are found in the
supplemental material of \cite{Colquhoun:2022atw}.  We have two
targets of physics to study: one is $T=0$ 2+1 flavor focusing on
B-physics, and the other is finite $T$ focusing chiral symmetry (2 and
2+1 flavors) and investigation of the Columbia plot (3 and 2+1
flavors).

The $T=0$ ensembles are $32^3\times 64$ -- $64^3 \times 128$ lattices
with pion mass $M_\pi$ between 230 -- 500~MeV, and the lattice cutoff
$a^{-1}$ is 2.5, 3.6 and 4.5~GeV \cite{Colquhoun:2022atw}.  Together
with 3 flavor ensembles with smaller volumes, we have more than 16
ensembles with 1.5k configurations.

Most of the finite-$T$ 2+1 flavor sets are generated along the line of
constant physics at $m_l=m_s^{\text{phys}}/27.4$ and
$m_l=m_s^{\text{phys}}/10$ with $N_T=12,~16$, and the remaining
finite-$T$ ensembles (2, 2+1 and 3 flavors) are generated with the
fixed $T$ approach.  Some up-to-date details are found in
\cite{Aoki:2021qws} (2 flavor), \cite{Zhang:2024ldl} (3 flavor), and
\cite{JLQCD:2024xey} (2+1 flavor).  The recent configurations are
generated mainly on the Fugaku supercomputer using Grid
\cite{Boyle:2016lbp}.  We employ the Hybrid Monte Carlo (HMC) and
Rational Hybrid Monte Carlo algorithms.  For the three flavor systems,
we apply the same 2+1 flavor algorithm but with degenerate masses.
The generated configurations are stored in a grid file system, the
Japan Lattice Data Grid (JLDG). We are in close contact with ILDG
members and plan to start uploading ensembles.  The number of
configurations and storage quoted in Table~\ref{tab:summary} assumes
storing every 100~MD trajectories, which can be revised when uploading
each ensemble.

\subsection{MILC}
The MILC Collaboration has been creating ensembles of gauge
configurations for decades.  Initial calculations used two dynamical
flavors with naive staggered quarks.  Second generation efforts used
dynamical up, down, and strange quarks with the asqtad action,
culminating with a review in Ref.~\cite{MILC:2009mpl}.  Our third
generation calculations use the highly improved staggered quark (HISQ)
action \cite{Follana:2006rc}.  They contain dynamical up, down,
strange, and charm quarks.  For most of the ensembles, up and down are
degenerate.  The charm quark is set near its physical value.  For
several ensembles, the light quarks are near their average physical
value.  There are also ensembles with $m_l= 0.1,$ or $0.2 m_s$, where
$m_l$ ($m_s$) is the light (strange) quark mass.  We have generated a
number of ensembles with $m_s$ less than its physical value to explore
low energy constants and the chiral Lagrangian.  Lattice spacings are
in the range of $[0.15, 0.03]$ fm.  In a few cases, multiple volumes
are available.  Details about the generation of configurations can be
found in Refs.~\cite{MILC:2010pul,MILC:2012znn,Bazavov:2017lyh}.

The sharing policy for the MILC ensembles is available on
GitHub~\cite{MILC:policy}. On that web page can be found links to 1)
the sharing policy, 2) a Google Sheet detailing freely available
ensembles, and 3) a document summarizing which papers to cite for the
use of each ensemble.  Anyone wishing to use an ensemble that is not
listed in the Google Sheet, but has been used in a publication or
noted in a talk, is welcome to contact a member of the Fermilab
Lattice or MILC Collaborations to inquire as to whether the ensemble
can be made available for a specific project.  Many configurations are
available on USQCD resources, making access relatively easy for USQCD
members.  The ILDG is not operational in the US, but should it become
so, we would make an effort to use it.  We have assisted transfer of
configurations to other researchers both within and outside the US.

\subsection{CLS}
The CLS (Coordinated Lattice Simulations) effort uses the openQCD code
\cite{openqcd} to generate ensembles
\cite{Bruno:2014jqa,Mohler:2017wnb} with \Nf{2}{1} non-perturbatively
$\mathrm{O}(a)$-improved Wilson quarks and tree-level improved
Symanzik glue, mostly with open boundary conditions in time to avoid
topological freezing \cite{Schaefer:2010hu,Luscher:2012av}, but also
with (anti-)periodic boundary conditions in time (on some ensembles at
$a\gtrsim 0.06\,$fm), at six fine lattice spacings
$a\in[0.039,0.1]$\,fm and quark masses from the symmetric to the
physical point on three chiral trajectories
($\mathop{\mathrm{Tr}}[M]=\mathrm{const.}$,
$m_s\approx\mathrm{const.}$, $m_s=m_l$) in large volumes satisfying
$M_\pi L\ge4$ throughout, with statistics typically $\gtrsim 2,000$
MDU.

Due to the algorithm used, two reweighting factors are needed to use
CLS ensembles: one to correct for the twisted-mass stabilization of
the light quarks, and one to correct for the rational approximation to
$\sqrt{D^{\dagger}D}$ for the strange quark.  In the latter case,
$\det D<0$ can occur \cite{Mohler:2020txx}, so that one also needs to
correct for the wrong sign of the reweighting factor; fortunately, the
fraction of configurations with a negative reweighting factor is very
small (or zero) for most ensembles.

At the time of Lattice 2024, there were 149,766 configurations (1.384
PB) stored on tape in the openQCD (non-ILDG) data format.  Metadata
regarding data provenance, simulation setup, HMC stability, and data
integrity are collected automatically via automated scripts, while
reweighting factors and determinant minus signs are measured
separately.  A first batch of ensembles has been successfully uploaded
to ILDG; several more ensembles will follow soon, and the remainder
will follow after an embargo period.  The XML metadata are generated
automatically by extraction from the existing database; the (signed)
reweighting factors are included in the Config XML.

\subsection{PACS}
In the past years, the PACS Collaboration has generated 2+1 flavor QCD
configurations on very large lattices close to the physical point
employing the stout-smeared $\mathcal{O}(a)$-improved Wilson-clover
quark action and Iwasaki gauge action. We use the stout smearing with
six iterations.  The improvement coefficient for the clover term is
non-perturbatively determined using the Schr\"odinger functional
scheme. These gauge configurations, which keep the space-time volumes
larger than (10 fm)$^4$, are called ``PACS10'' configurations.  We
have finished generating three gauge ensembles of (lattice spacing,
lattice size)=(0.085~fm, 128$^4$)~\cite{Ishikawa:2018jee}, (0.064~fm,
160$^4$)~\cite{Shintani:2019wai}, and (0.041~fm, 256$^4$), labeled as
PACS10/L128, PACS10/L160, PACS10/L256, respectively.  Using those
PACS10 configurations, we have studied the hadron
spectrum~\cite{Ishikawa:2018jee,PACS:2019ofv}, nucleon
structure~\cite{Shintani:2018ozy,Tsuji:2022ric,Tsuji:2023llh,Tsuji:2023qcr},
and physics beyond the standard
model~\cite{Shintani:2019wai,PACS:2019hxd,Ishikawa:2022ulx,Yamazaki:2023swq}.
For more precise determinations of physical quantities, we have
started to generate the PACS10$_c$ configurations, which are 2+1+1
flavor QCD configurations satisfying the PACS10 conditions.  In the
configuration generation, the degenerate up- and down-quarks are
simulated with the domain-decomposed HMC
algorithm~\cite{Luscher:2003vf} and the strange quark with the
rational HMC algorithm~\cite{Clark:2006fx}.  The up- and down-quark
determinant is calculated by separating into UV and IR parts, with
mass-preconditioning~\cite{Hasenbusch:2001ne} for the IR part,
even-odd preconditioning, and mixed precision nested
BiCGStab~\cite{PACS-CS:2008bkb} with the aid of the chronological
inverter as a guess.  Many gauge ensembles generated by the PACS
Collaboration and its predecessors, the CP-PACS and PACS-CS
Collaborations, are publicly available~\cite{JLDG}.  We plan to make
the PACS10 and PACS10$_c$ ensembles (roughly 200~TB) public through
the new generation of ILDG via JLDG after some embargo time.  Details
of the data policies are under discussion within the collaboration.

\section{Summary}

\begin{table}[h]
  \caption{ Public: (2 = currently public, 1 = after an embargo period, 0 = no); ILDG: (N = no interest, I =
    interest, P = planned, U = already using); \#ens: Number of
    ensembles; \#cfg: Total number of configurations; storage: Total
    storage needed in TBytes.
    \label{tab:summary}
  }
  \centering
  \begin{tabular}{lccrrr}\hline\hline
    Collaboration                & Public & ILDG & \#ens & \#cfg   & Storage (TB) \\\hline
    CLQCD                        & 1      & I    & 20    & 10,000  & 100          \\
    Jlab/W\&M/LANL/MIT/Marseille & 0      & N    & 13    & 10,000  & 3,500        \\
    HotQCD                       & 2,1    & P    & 70    & 15,000,000        & 1,500             \\
    FASTSUM                      & 2,1    & I    & 37    & 37,000  & 65           \\
    TELOS                        & 1      & P    & 250   & 800,000 & 120          \\
    HAL QCD                      & 1      & P    &  1    &  8,000  & 400          \\
    TWEXT                        & 1      & P    & 80    & 70,000  & 80           \\
    QCDSF                        & 1      & P,U  & 24    & 20,000  & 55           \\
    OpenLat                      & 2      & U    & 12    & 40,000  & 500          \\
    RC$^\star$                   & 1      & P    & 14    & 28,000  & 60           \\
    ETMC                         & 1      & P,U  & 24    & 12,000  & 3,000        \\
    JLQCD                        & 1      & P    & 260   & 37,000  & 15           \\
    MILC                         & 1      & I    & 50    & 70,000  & 600          \\
    CLS                          & 1      & P,U  & 64    & 150,000 & 1,400        \\
    PACS                         & 2,1    & P,U  & 6     & 200     & 200          \\\hline\hline
  \end{tabular}
\end{table}

The contributions presented in this proceeding demonstrate the diverse
approaches being taken in lattice QCD simulations. The fermionic
actions used by the contributing groups include staggered quarks,
particularly using the HISQ action (MILC, and CLQCD), improved
Wilson quarks on isotropic lattices (CLQCD,
Jlab/W\&M/LANL/MIT/Marseille, PACS, CLS, and QCDSF), improved Wilson
quarks on anisotropic lattices (FASTSUM and CLQCD), twisted mass
fermions (ETMC and TWEXT), Domain Wall fermions (JLQCD),
and Stabilized Wilson Fermions (OpenLat).

The simulations span both zero and finite temperature studies, with
some collaborations focusing exclusively on $T > 0$, such as TWEXT,
others on $T = 0$, and some pursuing both regimes, such as FASTSUM and
MILC. Most simulations employ either \Nf{2}{1} or \Nf{2}{1}{1} sea
quark flavors, with all the more ensembles being tuned to their
physical values.

A key development reflected in this year's contributions is the
growing effort to generate ensembles at finer lattice spacings, with
several collaborations now reaching $a\simeq0.04$~fm or smaller. This
push towards the continuum limit is particularly important for
precision calculations involving heavy quarks and for controlling
discretization effects in general.

The summary of ensemble generation efforts and data management
characteristics is presented in Table~\ref{tab:summary}. The collected
data indicate that most collaborations plan to make their ensembles
publicly available after an embargo period, with strong interest in
utilizing ILDG infrastructure for data sharing. The total storage
requirements reported by the collaborations are expected to exceed
10~PB, highlighting the substantial data management challenges faced
by the community.

Several trends can be identified in the current ensemble generation
efforts, namely: i) Increased focus on physical or near-physical quark
masses; ii) Growing emphasis on large physical volumes, with some
collaborations targeting (10 fm)\textsuperscript{4} or larger; iii)
Development of new algorithmic approaches to address challenges in
ensemble generation; iv) Enhanced attention to metadata collection and
provenance tracking; and v) An interest in open science principles,
with most groups planning public release of their data

The challenge of maintaining long-term storage infrastructure and
providing reliable access to these valuable datasets remains a
critical concern for the community. While ILDG 2.0 allows for
standardizing data sharing, the responsibility for storage space and
persistent accessibility remains with regional facilities and
individual collaborations. This highlights the ongoing need for
sustainable mechanisms to support storage resources required for
lattice QCD calculations, similar to those currently available for
computation.

The recorded level of interest in ILDG participation demonstrated by
the contributing groups suggests there is an appreciation in the
community for data sharing, and for frameworks with well-defined
workflows that enable this. Thus, standardized data management and
sharing practices may help to ultimately enhance the accessibility and
reusability of lattice QCD ensembles.

\acknowledgments

We acknowledge the support of all members of the ILDG for making the
data session possible. Y.~A. and G.~K.  would also like to thank
Hubert Simma for his assistance during the organization of the session
and the authoring of these proceedings.

The work of E.~B. is part-funded by the UKRI Science and Technology
Facilities Council (STFC) Research Software Engineer Fellowship
EP/V052489/1, the UKRI Engineering and Physical Science Research
Council (EPSRC) ExCALIBUR programme ExaTEPP project EP/X017168/1, and
the STFC Consolidated Grant ST/T000813/1. R.~B. acknowledges support
from a Science Foundation Ireland Frontiers for the Future Project
award with grant number SFI-21/FFP-P/10186. K.~U.~C. is supported by
the Australian Research Council grant DP190100297, DP220103098 and
DP240102839. T.~D. acknowledges support from HPCI System Research
Project (Grants No. hp200130, hp210165, hp220174, hp230207, hp240213,
hp220066, hp230075, hp240157, hp210212, and hp220240), the JSPS
(Grants No. JP18H05407, JP18H05236, JP19K03879, JP21H05190,
JP23H05439), JST PRESTO Grant No. JPMJPR2113, ``Program for Promoting
Researches on the Supercomputer Fugaku'' (Grants No. JPMXP1020200105,
JPMXP1020230411). S.~G. acknowledges support from the Department of
Energy, Office of High Energy Physics, DE-SC0010120. G.~K. acknowledges
support from EXCELLENCE/0421/0195, co-financed by the European
Regional Development Fund and the Republic of Cyprus through the
Research and Innovation Foundation and the AQTIVATE a Marie
Sk\l{}odowska-Curie Doctoral Network GA~No.~101072344.
G.~P. acknowledges funding by the Deutsche Forschungsgemeinschaft
(DFG, German Research Foundation) - project number 460248186
(PUNCH4NFDI). The work of C.~S. is supported by the European Union's
Horizon 2020 research and innovation program under the Marie
Sklodowska-Curie Grant Agreement No. H2020-MSCAITN-2018-813942
(EuroPLEx), by The Deutsche Forschungsgemeinschaft (DFG, German
Research Foundation) - Project No. 315477589-TRR 211 and the
PUNCH4NFDI consortium supported by the Deutsche Forschungsgemeinschaft
(DFG, German Research Foundation). T.~Y.  acknowledges Grants-in-Aid
for Scientific Research from the Ministry of Education, Culture,
Sports, Science and Technology (No. 19H01892, 23H01195, 23K25891),
MEXT as ``Program for Promoting Researches on the Supercomputer
Fugaku'' (JPMXP1020230409) and support from computational resources
through the HPCI System Research Project (Project IDs: hp170022,
hp180051, hp180072, hp180126, hp190025, hp190081, hp200062, hp200167,
hp210112, hp220079, hp230199, hp240207). Y.-B.~Y. is supported in part
by NSFC grants No.~12293060, and~12435002.

The participating collaborations acknowledge the following HPC systems
and HPC sites for the generation of the gauge ensembles reported here,

\begin{itemize}[leftmargin=*, align=left, itemsep=1pt, topsep=0pt, parsep=0pt]
\item National Computer Infrastructure (NCI) National Facility in
  Canberra, supported by the Australian Commonwealth Government and
  the Pawsey Supercomputing Centre, supported by the Australian
  Government and the Government of Western Australia, Australia;
  
\item HPC Cluster of ITP-CAS, the Southern Nuclear Science
  Computing Center (SNSC) in Beijing and the Siyuan-1 cluster,
  supported by the Center for High Performance Computing at Shanghai
  Jiao Tong University, China;
  
\item LUMI-C and LUMI-G at CSC, Finland;
  
\item SuperMUC and SuperMUC-NG at the Leibniz Rechenzentrum (LRZ) in
  Garching, JUGENE, JUWELS, and JUWELS-Booster at the Jülich
  Supercomputing Centre (JSC) in J\"ulich, HAWK at
  Höchstleistungsrechenzentrum Stuttgart (HLRS), and the
  North-German Supercomputer Alliance (HLRN), Germany;

\item  Irène Joliot-Curie at Très Grand Centre de Calcul (TGCC) in
  Bruyères-le-Châtel, France;

\item Kay and Stokes at the Irish Centre for High-End Computing
  (ICHEC) in Galway, Ireland;

\item Leonardo, Marconi 100, and Marconi A2 at CINECA in Bologna,
  Italy;
  
\item Polarie and Grand Chariot at Hokkaido University, the IBM Blue
  Gene system at KEK, SQUID at Osaka University, Supercomputer
  Fugaku at RIKEN, Oakforest-PACS at University of Tokyo, and
  Wisteria at University of Tsukuba, Japan;

\item Cambridge Service for Data Driven Discovery (CSD3) in Cambridge,
  Tesseract at the DiRAC Extreme Scaling Service at the University of
  Edinburgh, DiRAC Data Intensive 2.5 \& 3 at the University of
  Leicester, and Sunbird of Supercomputing Wales, supported by the
  European Regional Development Fund via the Welsh Government, United
  Kingdom;
    
\item Cori, Edisson, and Perlmutter at the National Energy Research
  Scientific Computing Center (NERSC), California; Mira, Theta,
  Polaris at the Argonne Leadership Computing Center (ALCF), Illinois;
  Blue Waters and Delta at the National Center for Supercomputing
  Applications (NCSA), Illinois; Big Red 2, 2+, 3, 200 at Indiana
  University, Indiana; Computing at Los Alamos National Lab, New
  Mexico; Summit, Crusher, and Frontier at the Oak Ridge Leadership
  Computing Facility (OLCF), Tennessee; Stampede, Stampede2, and
  Frontera at the Texas Advanced Computing Center (TACC), Texas; and
  the computing facilities at Jefferson Lab, Virginia, United States.
\end{itemize}

\bibliographystyle{JHEP} \bibliography{refs}

\end{document}